\documentstyle[12pt,aaspp4]{article}
\begin{document}

\title{CTQ 839: Candidate for the Smallest Projected Separation Binary Quasar \altaffilmark{1}}

\author{
 Nicholas D. Morgan\altaffilmark{2},
 Greg Burley\altaffilmark{3},
 Edgardo Costa\altaffilmark{4},
 Jos\'e Maza\altaffilmark{4,5}
 S. E. Persson\altaffilmark{3},
 Maria Teresa Ruiz\altaffilmark{4},
 Paul L. Schechter\altaffilmark{2},
 Ian Thompson\altaffilmark{3},
 Joshua N. Winn\altaffilmark{2}
}

\altaffiltext{1}{Based on observations carried out at the Cerro Tololo
 Interamerican Observatory (CTIO), the Las Campanas Observatory (LCO), and 
 and the National Radio Astronomy Observatory (NRAO) Very Large Array (VLA).
 CTIO is part of the National Optical Astronomy Observatories,
 which are operated by the Association of Universities for Research in
 Astronomy, Inc., under cooperative agreement with the National Science
 Foundation.  The NRAO is a facility of the National Science Foundation 
 operated under cooperative agreement by Associated Universities, Inc.}

\altaffiltext{2}{Department of Physics, Massachusetts Institute of
 Technology, Cambridge MA 02139; ndmorgan@mit.edu, schech@achernar.mit.edu,
 jnwinn@mit.edu}

\altaffiltext{3}{Carnegie Observatories, 813 Santa Barbara Street,
 Pasadena, CA 91101; burley@ociw.edu, persson@ociw.edu, ian@ociw.edu}

\altaffiltext{4}{Departamento de Astronom\'{\i}a, Universidad de Chile,
 Casilla 36-D, Santiago, Chile; ecosta@das.uchile.cl, jose@das.uchile.cl, 
 mtruiz@das.uchile.cl}

\altaffiltext{5}{C\'atedra Presidencial de Ciencias 1996-1998}

\begin{abstract}

We report the discovery of the new double quasar CTQ 839.  This $B = 18.3$,
radio quiet quasar pair is separated by $2\farcs1$ in $BRI$ \& $H$ filters with
magnitude differences of $\Delta m_B = 2.5$, $\Delta m_R = \Delta m_I = 1.9$, 
and $\Delta m_H = 2.3$. Spectral observations reveal both components 
to be $z = 2.24$ quasars, with relative redshifts that agree at the $100 
\mbox{ km s}^{-1}$ level, but exhibit pronounced differences in the equivalent
widths of related emission features, as well as an enhancement of blue 
continuum flux in the brighter component as compared to the fainter 
component longward of the Ly $\alpha$ emission feature.  In general,
similar redshift double quasars can be the result of a physical
binary pair, or a single quasar multiply imaged by gravitational 
lensing.  Empirical PSF subtraction of $R$ and $H$ band images of CTQ 839 
reveal no indication of a lensing galaxy, and place a 
detection limit of $R = 22.5$ and $H = 17.4$ for a third 
component in the system.  For an Einstein-de Sitter cosmology and SIS model, 
the $R$ band detection limit constrains the characteristics of any lensing 
galaxy to $z_l \gtrsim 1$ with a corresponding luminosity of $L \gtrsim 5 
L_*$, while an analysis based on the redshift probability distribution for the 
lensing galaxy argues against the existence of a $z_l \gtrsim 1$ lens at the 
2$\sigma$ level. A similar analysis for a $\Lambda$ dominated cosmology, 
however, does not significantly constrain the existence of any lensing galaxy. 
The broadband flux differences, spectral dissimilarities, and failure to detect
a lensing galaxy make the lensing hypothesis for CTQ 839 unlikely.  The 
similar redshifts of the two components would then argue for a 
physical quasar binary.  At a projected separation of 8.3 $h^{-1}$ kpc ($
\Omega_m = 1$), CTQ 839 would be the smallest projected separation binary 
quasar currently known.  

\end{abstract}


\section{INTRODUCTION}

The discovery of similar redshift, small separation optical-optical ($O^2$) 
double quasars (pairs where both components are optically bright and 
radio faint; see Kochanek, Falco, and Mu\~noz (1999)) can yield a range of 
information on cosmological scales.  
Such systems are intensively investigated as gravitational lens candidates, 
and if confirmed, can yield measurements of the Hubble constant (Refsdal 1964) 
as well as statistical constraints on the cosmological constant (Kochanek 
1996).   If an $O^2$ pair is confirmed as a binary quasar, it can provide 
clues regarding the triggering of nuclear activity in galaxies (Osterbrock 
1993) as well as the evolution of early ($z > 2$) galaxy mergers (Barnes 1999).
The observed frequency of binary quasars are also important in understanding
gravitational lensing statistics; Kochanek, Falco, and Mu\~noz (1999) have 
recently used the observed paucity of $O^2R^2$ quasar pairs (pairs with
both components bright in optical and radio) as compared to the 
number of $O^2$ pairs to conclude that the majority of known wide 
separation quasar pairs must be binary quasars.  In this paper, we report the 
discovery of the new small separation $O^2$ quasar pair CTQ 839 and 
investigate the nature of the system as either a gravitational lens or 
binary quasar.

CTQ 839 (2$^{\scriptsize{\mbox{h}}}$ 52$^{\scriptsize\mbox{m}}$
57$\fs$86, -32$\arcdeg$ 49$\arcmin$ 8$\farcs$6, J2000.0) was originally 
identified as a $z=2.24$ quasar from the Cal\'an-Tololo Survey 
(CTS) (Maza {\it et al.} 1996).  The CTS is an objective prism survey 
conducted at Cerro Cal\'an using photographic plates obtained at the
Cerro Tololo Interamerican Observatory (CTIO) and is aimed at 
discovering quasars and emission-line galaxies in the southern hemisphere.  To 
date, the CTS has identified $\sim$ 1000 southern hemisphere quasars as well as
two confirmed gravitational lenses: CTQ 286 (Claeskens {\it et al.} 1996) and 
CTQ 414 (Morgan {\it et al.} 1999).  During November 1998, approximately 100 
CTS quasars were observed with the 1.5 m telescope at CTIO as part of a five 
night observing program to discover new gravitationally lensed quasars.  This
particular run has yielded one definite lensed system, the complex 
gravitational lens HE 0230-2130 (Wisotzki {\it et al.} 1999), in addition to 
the double quasar CTQ 839 presented here.  Optical images of CTQ 839 
immediately revealed two components in the system, with a separation of 
2\farcs1 evident in all observed filters.  We describe these observations, as 
well as followup $R$ and $H$ band observations conducted at the Las Campanas 
Observatory (LCO), in \S 2.  In \S 3, we present our analysis of the 
quasar components from spectra taken at CTIO, while \S 4 
presents radio observations taken with the Very Large Array (VLA) in July of 
1999.  In \S 5 we discuss lens modeling and interpretation of the system, 
while \S 6 summarizes our findings and conclusions for CTQ 839.

\section{OBSERVATIONS AND REDUCTION}

\subsection{Initial Optical Imaging}

Initial optical observations of CTQ 839 were taken with the CTIO 1.5 m 
telescope by two of us (N.D.M. and P.L.S.) on the nights of 1998 November 
12 and 15.  The SITe 2048 \#6 CCD camera was used, although only 
the central 1536 $\times$ 1536 of the array was read out.  The telescope was 
operated at an $f$ ratio of $f/13.5$, providing a field of view of 6.1 arcmin 
square and a scale of 0$\farcs$2407 per pixel.  The gain and read noise of the 
detector were 2.9 e$^-$/ADU and 4.0 e$^-$, respectively.  Multiple 300 s 
exposures of CTQ 839 were taken in Johnson $B$ and $R$ and Cousins $I$ filters 
with FWHM seeing conditions ranging from 0$\farcs$9 to 1$\farcs$1.  Multiple 
$BVRI$ exposures of the Landolt standard field Rubin 149 (Landolt 1992) 
were also taken on the 15$^{\it{th}}$ for use in photometric calibration.  
Table 1 presents a log of the CTIO observations.  Figure 1 shows a 6 
arcmin square exposure of CTQ 839 and nearby stars from one of the $I$ band 
frames.

The CCD frames were bias subtracted, trimmed, and flatfield corrected using 
the Vista reduction program.  The flatfield frames consisted of twilight 
exposures taken on multiple nights of the CTIO observing run, and were cleaned
of cosmic rays using ``autoclean'', a program written and 
kindly supplied by J. Tonry.  As noted in \S 1, images of CTQ 839 
are separable into two components in all filters (see Figure 2).  The double 
images were therefore fit with two empirical point spread functions (PSFs) 
using a variant of the program DoPHOT (Schechter {\it et al.} 1993), designed 
to deal with close, point-like and extended objects (Schechter and Moore 1993).
Star \# 5 identified in Figure 1 provided the empirical PSF.  Results
for the relative positions and apparent magnitudes of the brighter and 
fainter components (denoted by A and B, respectively) are presented in 
Table 2.  Here we present results from simultaneous fitting of magnitudes 
and relative positions of the two components.  Magnitude solutions using fixed
separations differ by $< 0.01$ mag in all filters.  One of the $B$ band frames
(\#54), which registered a cosmic ray detection $\sim 1\arcsec$ north of 
component B, was omitted from the analysis.

It can be seen that the A:B flux ratio exhibits a rather strong dependence 
with filter, dropping from -2.56 mag in $B$ band down to -1.85 mag
at $R$ and $I$ wavelengths.  The $B$ band separation of $2\farcs$064 is
also $\sim$ 0$\farcs$03 smaller than the separations found at $R$ and $I$ 
wavelengths of 2$\farcs$098 and 2$\farcs$092, respectively.  The reason for
this difference in image separations at blue and red wavelengths is not 
immediately clear to us, although we have ruled out variations in the CCD 
scale as a possible source.  Although the separations are consistent with 
gravitational lensing, the wavelength dependent flux ratio would require 
strong reddening and/or microlensing of the quasar's light to conform with 
the lensing hypothesis.  PSF subtraction of components A and B using stacked 
images for each filter showed no indication of a third component in the 
system, although deeper and redder searches for possible signs of a lensing 
galaxy are described in the following subsection.

The observations reported above used the 1.5 m telescope at CTIO operating at a
focal length of $f/13.5$.  The 1.5 m is a Ritchey-Chr\'etien telescope, 
designed to be free of comatic aberration at $f/7.5$, but not at $f/13.5$.   
Observations were carried out at $f/13.5$ in order to make use of the smaller 
pixel scale at that $f$ ratio.  The observations reported above therefore 
suffer from coma, which introduces an off-axis distortion in the 
shape of the PSF across the CCD chip.  This effect grows with increasing 
distance from the center of the chip, and, given the relatively good seeing 
conditions during the observing run, can cause PSF magnitudes to
systematically underestimate corresponding aperture magnitudes by as much
as 0.1 mag for peripheral stars.  In performing the PSF analysis described 
above, we were careful to choose a PSF star as close as possible
to CTQ 839 ($\sim$ 30 arcseconds away) in order to minimize the effects of
coma.  The resulting PSF and aperture magnitudes for the combined 
flux from components A and B agree to within 0.035 mag in $B$, and better 
than 0.010 mag in $R$ and $I$. 

For use with future observations, aperture magnitudes were determined for 
8 field stars within a 4 arcminute radius from the target quasar.  
An aperture diameter of 9\farcs6 was used.  Observations were calibrated 
using the Rubin 149 standard field (Landolt 1992) mentioned above, with
extinction coefficients taken from the 1990 CTIO Facilities Manual 
(\begin{math} k_B = 0.22, k_V = 0.11, k_R = 0.08, k_I = 0.04 \end{math}).
These results are presented in Table 3, along with corresponding astrometric 
solutions for the selected reference stars. 

\subsection{Follow-up Optical and Infrared Imaging}

In an effort to further probe the system for the possible presence of a 
lensing galaxy, follow-up $R$ and $H$ band observations were carried out 
within a few months of the original observations.  On 23 December 1998, a 
series of six 10 minute $R$ band exposures of CTQ 839 were taken by one of us
(E.C.) with the du Pont 2.5 m telescope at LCO.  The Tek \#5 detector set in 
the \#3 gain position was employed, providing a gain of 3.0 e$^-$/ADU, a 
readnoise of 7.0 e$^-$, and scale of 0$\farcs$2604 per pixel.  Seeing 
conditions for the series of observations were slightly better than those 
taken at CTIO, with an average FWHM of 0$\farcs$95.  After bias-subtraction 
and flatfielding, the images were co-added using integer pixel shifts.  The 
resulting stacked image was then reduced in the same manner as described 
above.  Given the longer exposure times as compared to the CTIO data, star
\# 5 became saturated on the CCD detector and a new star (\# 4 in Figure 1) 
provided the empirical PSF.

Results from PSF analysis yield an A:B flux ratio of 5.64 and a separation
of 2$\farcs$092 at a PA of 160\fdg8 E of N. Both of these results compare well 
with the CTIO $R$ band solutions presented in Table 2.  In the two upper
panels of Figure 2, we show an excised portion from the stacked image of 
CTQ 839 (left), along with residuals after PSF subtraction (right).  (The 
bottom two panels show $H$ band observations taken two months later; see the 
following subsection).  In the residual panels, tick marks indicate 
the centroid locations of components A and B as determined from the PSF fits. 
The orientation of the images are the same in all panels of the figure.   

When ground-based observations of close separation, doubly lensed systems
are fit with two PSFs, a characteristic residual pattern emerges after
PSF subtraction.  These patterns arise from using only two PSFs to model the 
light from both quasar images as well as the lensing galaxy, and consist of 
undulating regions of positive and negative residuals.  For instance, one
typical residual pattern consists of a ``divot-bump-divot'' undulation, as 
seen by Schechter {\it et al.} (1998) and Morgan {\it et al.} (1999).  
These types of patterns are not present in the upper right 
panel of Figure 2, and therefore argues against the presence of 
any significant third component in the system.  (The residuals located around 
the center of component A, which do not show significant structure, are 
likely due to imperfections in component A's fit).  In order to place a 
magnitude limit on this null result, we inserted a series of gaussian 
profiles of varying magnitudes into the stacked $R$ band image and investigated
the residual pattern that emerged after fitting each system with two PSFs.  
The position of the gaussian profile was dictated by the singular isothermal 
sphere (SIS) model, that is, the center of the gaussian was placed collinear 
with the centroids of components A and B, with the ratio of relative 
separations from the two components given by the LCO $R$ band flux ratio.  The 
FWHM of the profile was dictated by the average seeing conditions for the LCO 
run.  The profile's magnitude, starting at the same brightness as component B, 
was successively dimmed by 0.1 mag increments until the characteristic 
residual pattern was no longer unmistakable.  We conclude that we would have 
confidently detected any third component in the system brighter than 
$R = 22.5$ at the expected position for a lensing galaxy. 

Following the null result in $R$ band, infrared observations of CTQ 839 in 
$H$ band (1.65 microns) were carried out by two of us (G.B. and I.T.) at LCO 
on 1999 February 5.  The IRCAM infrared camera (Persson {\it et al.} 1992) 
mounted on the du Pont 2.5 m telescope was used at a scale of $0\farcs3478$ 
per pixel.  The total integration time on CTQ 839 was 3250 seconds, 
divided into 13 individually dithered frames of $5 \times 50$ s  each.  The 
reduction procedures were carried out by one of us (S.E.P.) and followed 
closely those described in Persson {\it et al.} (1998).  The 13 fully 
processed frames were combined into a final stacked image, from which 
photometry was obtained.  Given the narrow (256 $\times$ 256) field of view
of the detector, only star \#5 as shown in Figure 1 was available to provide 
the empirical PSF.  This star is rather bright at $H = 13.4$, and 
as a consequence, the central pixel value for the object required a linearity 
correction at the $5\%$ level.  The residual error associated with this
correction affects aperture magnitudes for star \#5 no larger than 0.01 mag, 
which is not critical for the analysis that follows.

Results from empirical PSF analysis yield an $H$ band A:B flux ratio of 8.42, 
which is 1.5 times larger than the $R$ and $I$ band results.  An analysis 
using an analytical PSF, as described in Schechter {\it et al.} (1993),
yields an A:B flux ratio of 8.30.  The $m_A - m_B$ magnitude differences
between the empirical and analytical models therefore differ on the $0.02$ 
mag level.  The separation between the two components as determined from 
empirical PSF fitting was 2$\farcs$101 $\pm$ 0$\farcs$020, which agrees with 
the separations found at $R$ and $I$ wavelengths.  An excised portion of the 
$H$ band observation centered CTQ 839, along with residuals after fitting with 
two empirical PSFs, are shown in the bottom two panels of Figure 2.  The 
residual image again shows no indication of a significant third component 
in the system.  The clustering of positive residuals at the centroid of A's
fit, which are of order $5\%$ of A's peak intensity, are consistent with
the imperfection in the empirical PSF template described above.  Using the 
identical procedure outlined for the LCO $R$ band data, we conclude that we 
would have confidently detected a third component at the expected position 
of a lensing galaxy brighter than $H = 17.4$.

\section{SPECTROSCOPY}

Spectra of both components of CTQ 839 were obtained on 1998 December 27 
by one of us (M.T.R.) with the 4.0 m telescope at CTIO.  The R-C Spectrograph 
together with the Blue Air Schmidt camera and Loral 3 K $\times$ 1 K CCD were 
used.  The wavelength scale for the observations was 1.205 \AA$\;$ per pixel, 
with a wavelength range from 3670 to 7210 \AA, and a long, 1$\arcsec$ wide 
slit; seeing conditions were 1$\farcs$3 FWHM.  With the slit orientation 
placed perpendicular to the component separation, one 360 s exposure centered 
on component A and two 1800 s exposures centered on component B were taken.  
Three spectrophotometric standard stars from Baldwin and Stone (1984) were
also observed during the night for flux calibration purposes.  All spectra 
were bias-subtracted and flatfield corrected using standard IRAF procedures.  
The observations of CTQ 839 were carried out close to the zenith, with 
airmasses ranging from 1.001 to 1.010, so differential lightlosses due to 
atmospheric refraction were not a problem.  However, with a separation 
distance smaller than twice the seeing disc, some contamination of the 
fainter component's spectra with light from component A was unavoidable.  This 
contamination, which we have estimated to be $\sim$5\% of B's raw spectra, is 
straightforward to compensate for assuming gaussian profiles and a knowledge 
of the seeing disc and slit characteristics.  In Figure 3, we show the 
spectra of component A and the decomposed average spectra of component B, 
along with the identification of prominent emission features.  

The spectra of A and B both show quasar emission profiles at similar redshifts.
Both components exhibit appropriately redshifted Lyman $\alpha$ $\lambda$1216, 
\ion{N}{5} $\lambda$1240, and \ion{C}{4} $\lambda$1549 emission features, 
while A also displays \ion{O}{1} $\lambda$1304, \ion{Si}{4} $\lambda$1397 
+ \ion{O}{4} $\lambda$1402, and \ion{C}{3]} $\lambda$1909 emission lines as 
well.  Table 4 lists the strongest emission features for both 
components, as well as redshift determinations based on gaussian fits to 
the peaks of the profiles.  The redshifts of both spectra are consistent with 
a $z = 2.24$ quasar.  A cross-correlation between the two spectra yields 
relative redshifts that agree at the $100 \mbox{ km s}^{-1}$ level.

The quotient of the two spectra, shown in Figure 4, shows strong evidence for 
differences in the equivalent widths of related emission 
features.  The prominent peaks present in the quotient spectrum, corresponding
to the \ion{C}{3]}, \ion{C}{4}, and Ly $\alpha$+\ion{N}{5} emission features, 
are consistent with A having progressively stronger emission lines with 
respect its continuum than B does as one moves from the red to blue 
wavelengths.  There is also an indication for a harder blue continuum in A 
than in B, longward of the Lyman $\alpha$ emission feature.  As discussed
further in \S 6, these spectral differences between the two components
are difficult to reconcile under the lensing hypothesis. 

\section{RADIO OBSERVATIONS}

In order to search for radio emission from CTQ 839, we first queried the
NRAO VLA Sky Survey (NVSS) (Condon {\it et al.} 1998) at the position of the 
brighter optical component.  The NVSS is a 1.4 GHz radio continuum survey of 
all the sky north of $-40^{\circ}$, carried out with the NRAO Very Large Array
(VLA) in its D configuration.  The FWHM resolution is 45 arcseconds and the 
quoted completeness limit is 2.5 mJy.  However, no radio source was found 
within 3 arcminutes of the optical position.

A deeper probe for radio emission from CTQ 839 was performed by one of us
(J.N.W.) on 1999 July 21 using the VLA.  The search was carried out with a
15 minute integration at 8.4 GHz, while the VLA was in 
the A configuration.  The FWHM of the synthesized 
beam was 0.5 arcseconds in the N/S direction and 0.2 arcseconds in the E/W 
direction.  No significant sources of radio flux were detected within 5 
arcseconds of the position of the brighter optical component of CTQ 839.  
The rms noise level in this field was 0.17 mJy per synthesized beam, so our 
observation rules out (at the 5$\sigma$ level) any sources of compact flux 
above 0.85 mJy.  We therefore classify CTQ 839 as an $O^2$ quasar pair.

\section{SIS MODEL AND INTERPRETATION}

If CTQ 839 is a gravitational lens system, then the failure to detect a third
component can place constraints on the characteristics of any lensing galaxy 
that may be present \footnote{In this 
section, we will continue to refer to the ``lensing galaxy'', although we 
realize its existence is by no means conclusive.}.  In the following section, 
we use a simple SIS model to describe the galaxy potential in order to predict
the lensing galaxy's luminosity as a function of distance.  Combined with the 
magnitude detection limits discussed in \S 2, we investigate the types of 
bounds that can be placed on the lens galaxy evolutionary type and 
redshift.

The SIS model is characterized by three parameters: two angular coordinates for
the center of the potential, as well as the associated line-of-sight velocity 
dispersion $\sigma$ which measures the depth of the potential well.  For
a SIS model, the velocity dispersion of the lensing potential is related to 
the image separation $\theta$ by
\begin{equation}
\frac{\sigma ^2}{c^2} = \frac{D_S}{D_{LS}}\frac{\theta}{8 \pi}
\end{equation} 
where $D_S$ and $D_{LS}$ are angular diameter distances from the observer
to the source and from the lens to the source, respectively, and $\theta$ 
is measured in radians (see, for example, Narayan and Bartelmann 1998).
We assume the galaxy's central velocity dispersion is related to its $B$ 
band luminosity $L$ via a Faber-Jackson relationship of the form
\begin{equation}
\frac{L}{L_*} = \left(\frac{\sigma}{\sigma_*}\right)^{\gamma},
\end{equation}
where, following Keeton, Kochanek, and Falco (1998), we adopt $\sigma_* =
220 \mbox{ km s}^{-1}$, $\gamma = 4.0$ for early-type galaxies, and 
$\sigma_* = 144 \mbox{ km s}^{-1}$, $\gamma = 2.6$ for late-type galaxies. 
$L_*$ corresponds to a $B$ band magnitude of $M_B^* = -19.7 + 5\log h$, where 
the Hubble constant has been parameterized by $H_o = 100h 
\mbox{ km s}^{-1} \mbox{ Mpc}^{-1}$.  For a given redshift $z_l$ of the 
lensing galaxy, we can then estimate its cosmological distance modulus via
\begin{equation}
m_{AB}(\lambda_{obs}) - M_{AB}(\lambda_{rest}) = 5 \log \frac{D_L}{10 \mbox{ pc}} + 7.5 \log (1+z{_l})
\end{equation}
where $D_L$ is the angular diameter distance of the lensing galaxy.  For the 
purpose of calculating $M_{AB}(\lambda_{rest})$, spectral energy 
distributions (SEDs) for both early- and late-type galaxies were 
obtained from Lilly (1997), which consisted of interpolation and extrapolation
of the SEDs presented by Coleman, Wu, and Weedman (1980).  The SEDs are then 
normalized to the Faber-Jackson luminosity at $4400(1+z)$ \AA, which yields 
the predicted AB magnitudes.  Transformations to the $BVRI$ system from the 
$AB$ magnitudes were performed by adding -0.110, 0.011, 0.199, and 0.456, 
respectively, to the $AB$ magnitudes (Fukugita, Shimasaku, and Ichikawa 1995).

We have calculated predicted $R$ band magnitudes for the lensing galaxy 
as a function of lensing redshift for both an $\Omega_m = 1, 
\Omega_{\Lambda} = 0$ Einstein-de Sitter universe and an $\Omega_m = 0.3, 
\Omega_{\Lambda} = 0.7$ open universe (See Figure 5).  For the Einstein-de 
Sitter cosmology, it can be seen that the late-type spiral model lies above 
({\it i.e.}, brighter than) the detection threshold by a full magnitude for 
the entire range of $z_l$, and makes it therefore an unlikely model for the 
lensing galaxy.  For the same cosmology, the early-type elliptical model 
requires $1.0 \lesssim z_l \lesssim 2.0$ for consistency with the detection 
limit found in \S 2.  At a redshift of $z_l = 1$, the elliptical
galaxy model is already rather luminous, with an intrinsic luminosity of 
$\sim 5 L_*$ (corresponding to a velocity dispersion of $325$ km s$^{-1}$).  
We can estimate the likelihood of finding a lensing galaxy within the above
redshift range using the procedures of Kochanek (1992).  Using the 
critical lens radius of $r = 1\farcs045$ for CTQ 839, we compute a median 
redshift for the lensing galaxy of $z = 0.46$, with a $2\sigma$ probability
interval of $0.11 \leq z \leq 0.93$.  Thus, under the lensing 
hypothesis, the lensing galaxy ought to have been seen in an Einstein-de 
Sitter cosmology in more than 95\% of such cases.

The constraints on the existence of the lensing galaxy are far less stringent 
for the $\Lambda$ dominated cosmology.  For the $\Omega_m = 0.3, 
\Omega_{\Lambda} = 0.7$ universe, consistency with the $R$ band detection 
limit from \S 2 requires $0.7 \lesssim z_l \lesssim 2.1$.  The median 
redshift is found to be $z = 0.57$ with a $2\sigma$ probability range of 
$0.16 \leq z \leq 1.07$, which does not significantly constrain the 
redshift of the lensing galaxy.  Thus while the existence of the lensing
galaxy is highly unlikely in an Einstein-de Sitter universe, the
$\Lambda$ dominated model cannot argue for or against the lensing
hypothesis.

\section{SUMMARY AND CONCLUSIONS}

Although it may be attractive to explain CTQ 839 as a gravitational lens,
there are clearly a number of characteristics of the system that
make the lensing hypothesis less than convincing.  For example, 
while the observed image separation of $2\farcs1$ is typical of known double 
gravitational lens systems, the broadband A:B flux ratios (10.4:1 in $B$, 
5.5:1 in $R$ and $I$, and 8.4:1 in $H$) exhibit a rather 
large variation with wavelength.  Since the detection limit for a third 
component in the system is $\sim$ 22.5 in $R$, the 
smaller $R$ and $I$ band flux ratios is likely not flux augmentation of 
component B from an intervening galaxy.   Also, if this was the case, the 
SEDs of early-type galaxies would predict an even smaller flux ratio in $H$ 
band, which is not observed.  Extinction of component B's light by a line of 
sight absorber is a possible explanation, although such an absorber would 
have to preferentially absorb more flux at $B$ and $H$ wavelengths and less 
so at $R$.

Microlensing of quasar light by an intervening galaxy remains a 
possible explanation for the observed differences in flux ratios, although the
situation is highly contrived.  First, we note that the quotient spectra 
shown in Figure 4 exhibits an enhancement of blue continuum flux 
in component A as compared to component B shortward of the $\lambda_{obs} 
\sim 5500 $\AA$\;$ mark, which is consistent with microlensing of component 
A's light by stars in an intervening galaxy (Kayser {\it et al.} 1986).  Such 
an effect has already been observed in at least one confirmed gravitational 
lens, HE 1104-1805 (Wisotzki {\it et al.} 1993).  However, the observed 
differences in the line strengths of respective emission features for the
two components are difficult to reconcile under the microlensing scenario.  
The line strengths of the emission features, which are thought to arise from 
a region roughly an order of magnitude larger than the continuum emitting 
region, ought not to be strongly affected by microlensing.

We therefore conclude that CTQ 839 is unlikely to be a gravitationally lensed 
system.  The broadband flux differences, spectral dissimilarities, and failure
to detect a lensing galaxy all argue against (although do not explicitly rule 
out) the gravitational lensing explanation for CTQ 839.  If CTQ 839 is not 
a lens, it must be two separate quasars.  The nearly identical redshifts 
derived from the spectra of the two components would then argue for a 
physical binary system.  At a separation of 2\farcs1 and a redshift of 
$z = 2.24$, the projected separation of the system is 8.3 $h^{-1}$ kpc 
($\Omega_m = 1$), which would make CTQ 839 the smallest projected separation 
binary quasar currently known (Kochanek {\it et al.} 1998).

\acknowledgements 

N.D.M. and P.L.S. gratefully acknowledge the support of the U.S. 
National Science Foundation through grant AST96-16866.  J.N.W. thanks the 
Fannie and John Hertz Foundation for financial support.  J.M. thanks 
FONDECYT, Chile, for support through grant 1980172.  M.T.R. acknowledges 
partial support from FONDEYCT, Chile, through grant 19890659 and a C\'atedral
Presidencial (1996).

\clearpage
\figcaption[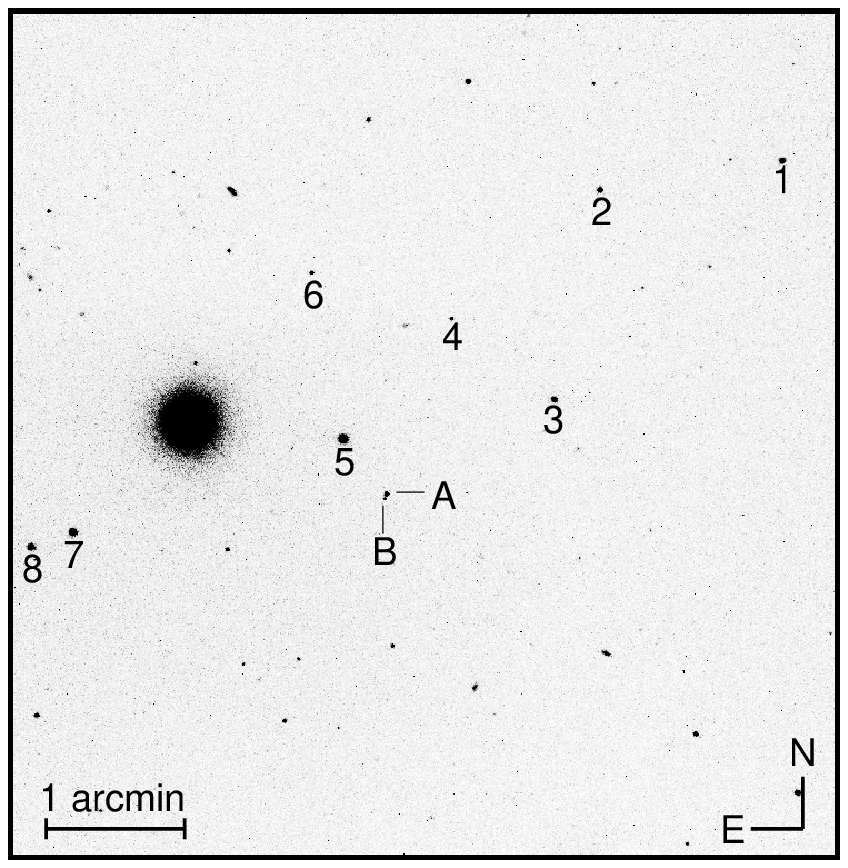] 
        { a) A 300 s {\em I} band image of CTQ 839 and the surrounding field 
taken with the 1.5 m telescope at CTIO.  The quasar components are labeled
by A (brighter) and B (fainter).  Photometric and astrometric solutions for 
stars labeled 1 through 8 are presented in Table 3.  North is up, and east is 
to the left.  The scale is shown in the bottom left of the figure. 
\label{fig1}}

\figcaption[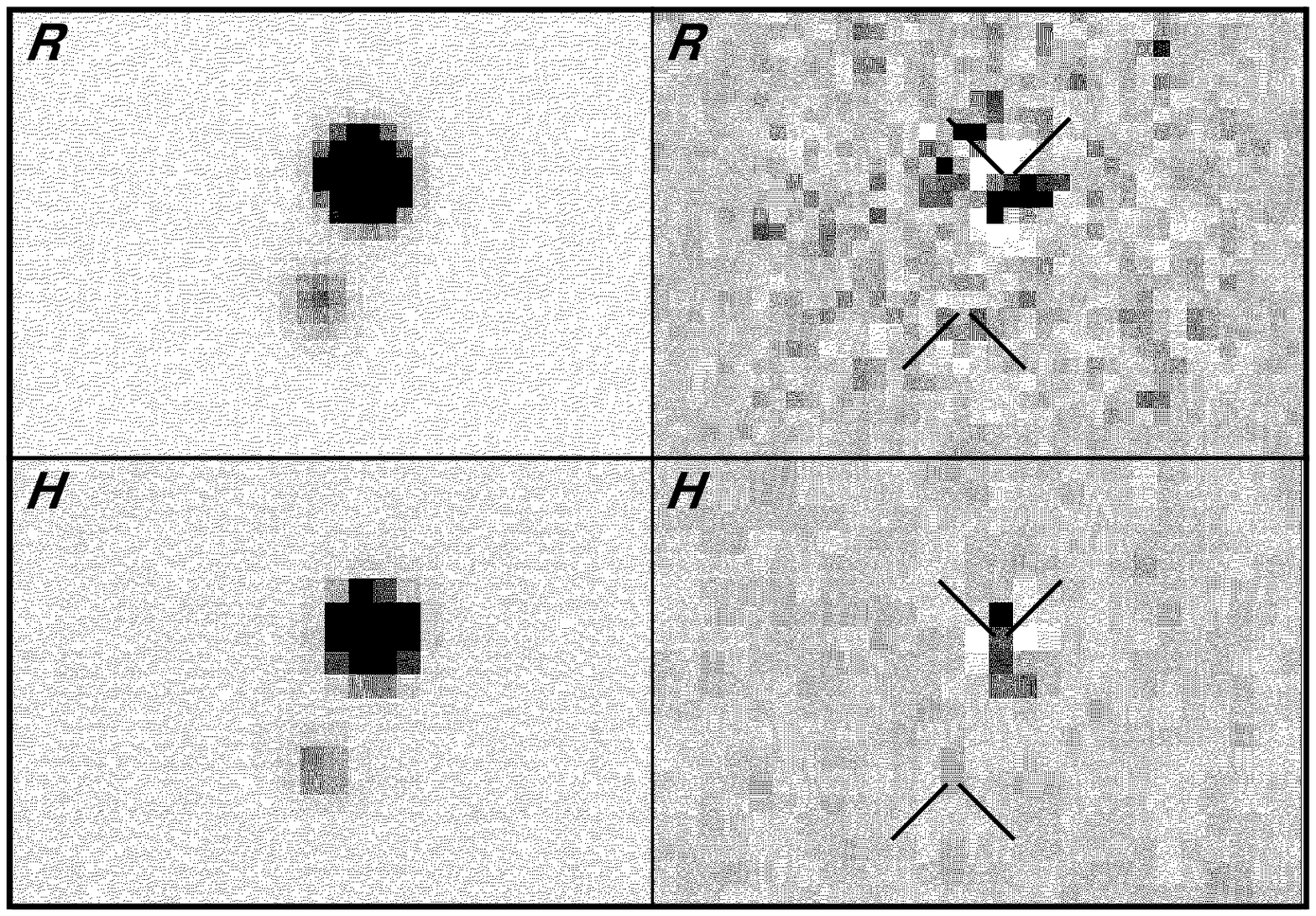]
	{ Excised CCD subrasters centered on CTQ 839 (left) along with 
residuals after empirical PSF subtractions (right).  The top two panels show 
the stacked $R$ band data taken at LCO (summed to $\sim$ 60 min), while the 
bottom two panels show the $\sim$ 54 min exposure in $H$ band.  The tickmarks
in the two residual panels mark the centroid location of components A and B
as determined from empirical PSF fitting.  The scale and orientation is 
identical on all panels, with North up and East to the left.  Saturation 
levels for the residual panels are $\pm 10\sigma$ where $\sigma$ is the 
respective sky noise for each filter, with gray scale intervals every 
$1\sigma$.  The peak of component A in $R$ ($H$) band is $\sim1840\sigma$ 
($\sim250\sigma$). \label{fig2}} 

\figcaption[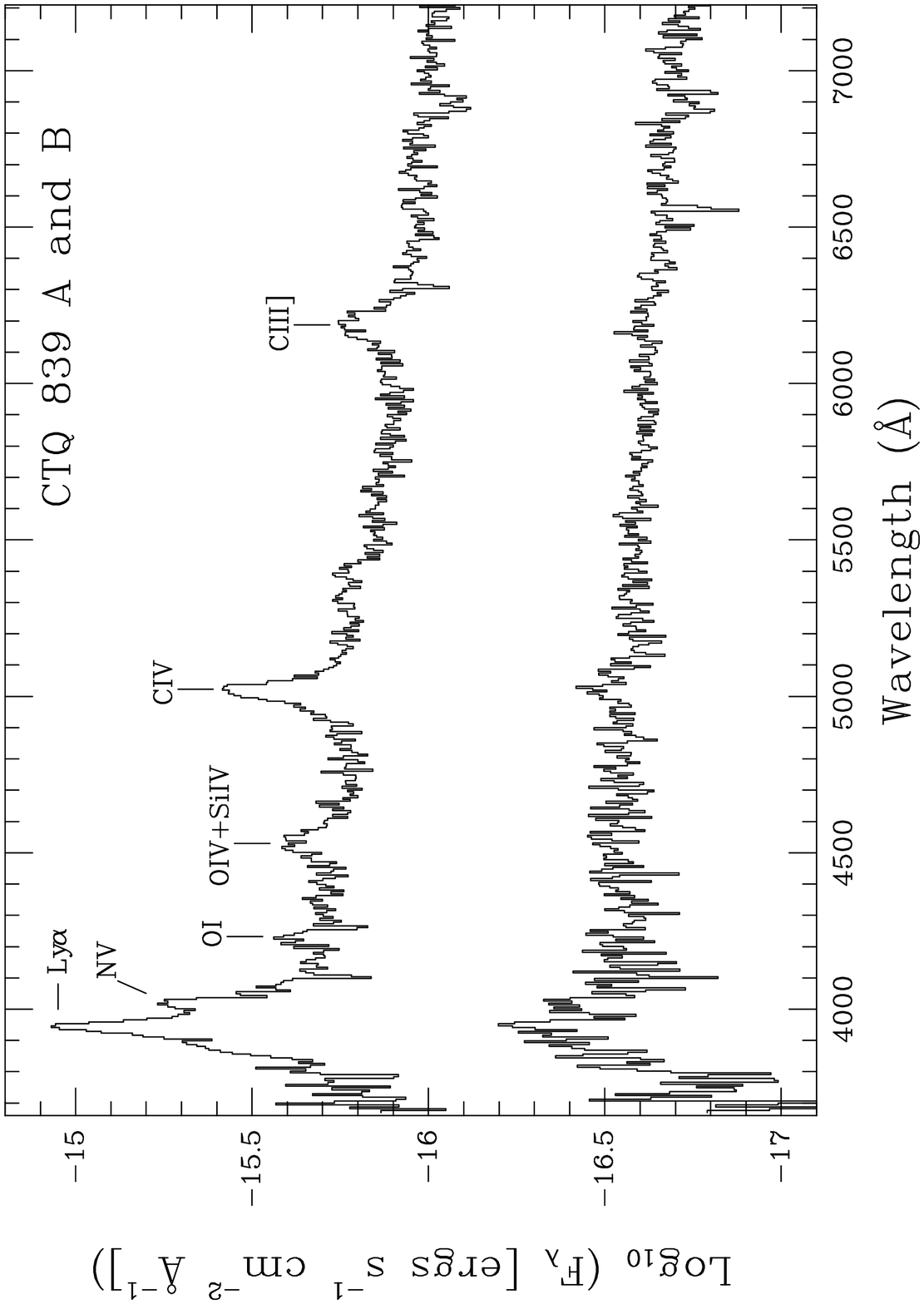] 
	{ Spectra of the brighter (A) and fainter (B) components of CTQ 839
taken with the 4.0 m at CTIO.  Both spectra have been binned by 6 \AA.
\label{fig3}}

\figcaption[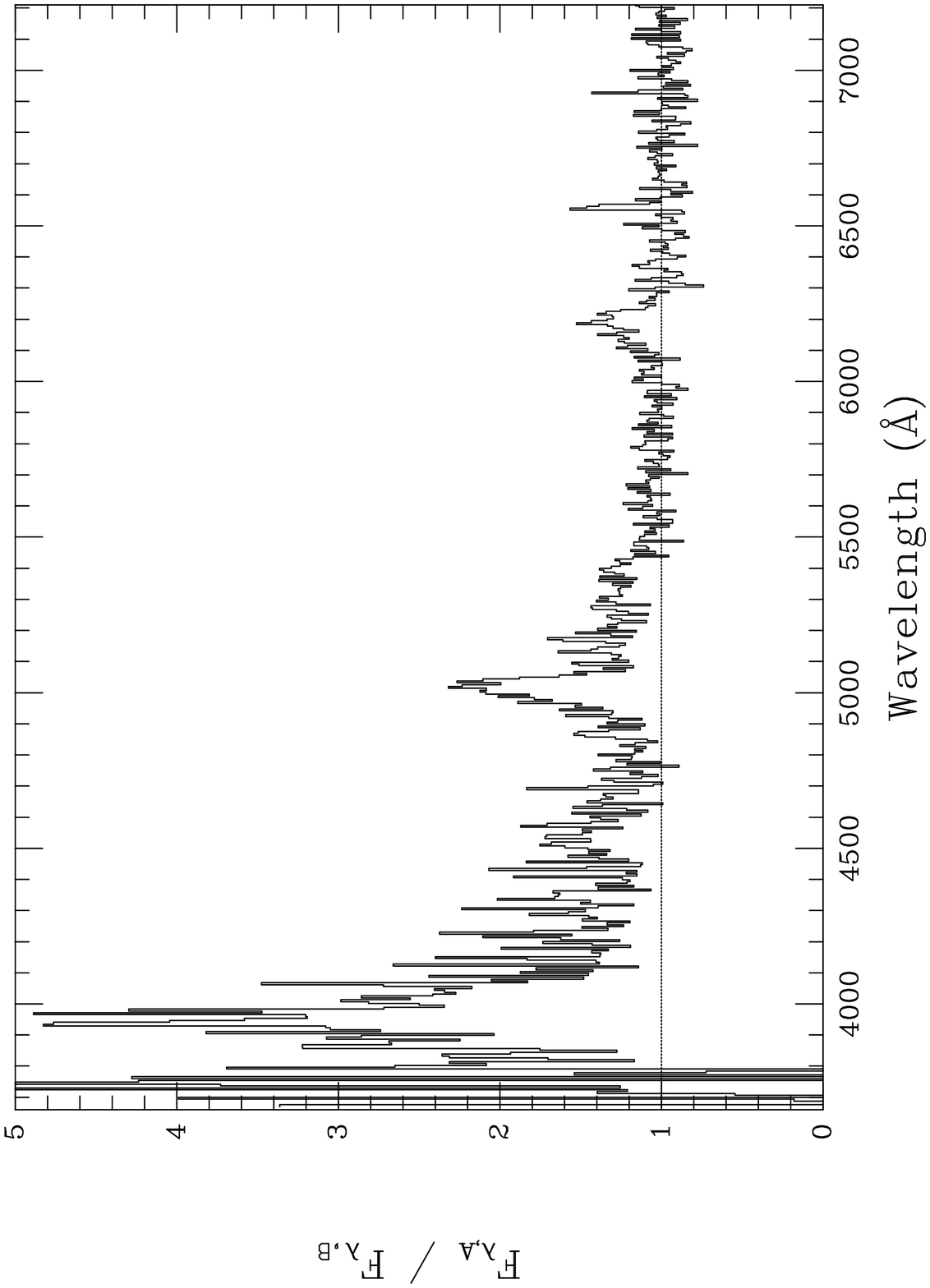] 
	{ Quotient spectra for CTQ 839, after normalizing at 6725 \AA. Bin
size is the same as in Figure 3. \label{fig4}}

\figcaption[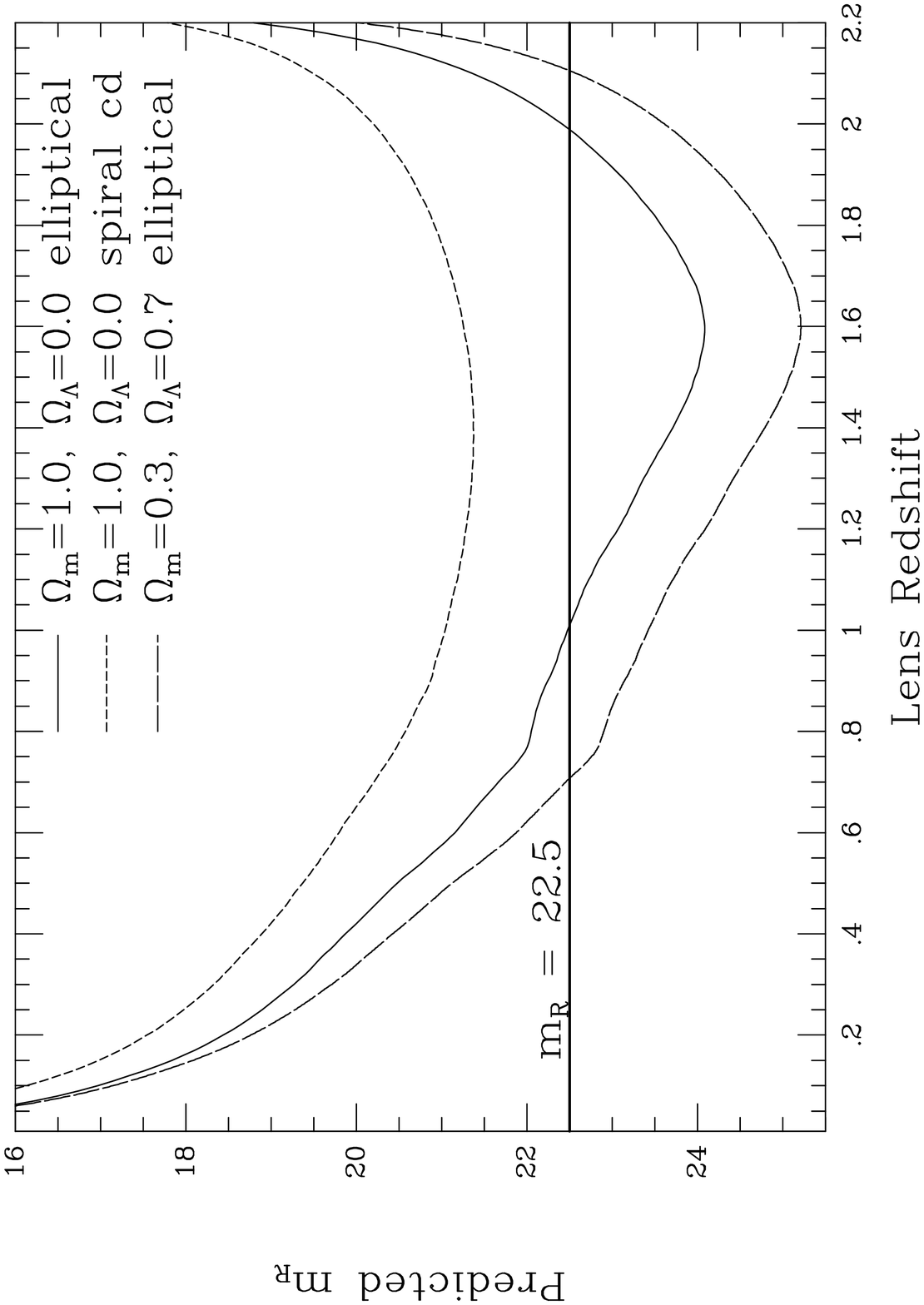] 
	{ Predicted $R$ band magnitudes for the lensing galaxy 
as a function of the lens redshift.  Two separate cosmologies are
considered.  The solid and short dashed curves show predictions for 
an early-type elliptical and late-type spiral, respectively, in an 
$\Omega_m=1, \Omega_{\Lambda}=0$ universe.  The long dashed curve 
shows predictions for an early-type elliptical in an $\Omega_m=0.3, 
\Omega_{\Lambda}=0.7$ universe.  The R band detection limit for a third 
component in the system is indicated by the heavy horizontal line.
\label{fig5}}

\clearpage




 

\makeatletter
\def\jnl@aj{AJ}
\ifx\revtex@jnl\jnl@aj\let\tablebreak=\nl\fi
\makeatother


\begin{deluxetable}{crcc}
\tablecaption{Log of CTIO Observations (Nov. 1998)
\label{TABLE1}}
\tablenum{1}
\tablewidth{0pt}
\tablehead{
\colhead {Frame \#} &
\colhead {Time (UT)} &
\colhead {Filter} &
\colhead {FWHM (\arcsec)}}
\startdata
052 & 98 Nov 12 05:49 & R & 0.88 \nl
054 &           06:00 & B & 0.99 \nl
346 & 98 Nov 15 05:11 & I & 1.00 \nl
347 &           05:17 & I & 1.03 \nl
348 &           05:24 & I & 1.02 \nl
349 &           05:30 & B & 1.09 \nl
350 &           05:35 & B & 1.04 \nl
351 &           05:41 & B & 0.99 \nl
352 &           05:48 & R & 0.96 \nl
\enddata
\end{deluxetable}


\clearpage



 

\makeatletter
\def\jnl@aj{AJ}
\ifx\revtex@jnl\jnl@aj\let\tablebreak=\nl\fi
\makeatother


\begin{deluxetable}{cccccc}
\tablecaption{Absolute Photometry and Relative Astrometry for CTQ 839
\label{TABLE2}}
\tablenum{2}
\tablewidth{0pt}
\tablehead{
\colhead {Filter} &
\colhead {$N_{im}$} &
\colhead {$\Delta$ RA ($\arcsec$)} &
\colhead {$\Delta$ Dec ($\arcsec$)} & 
\colhead {$m_A$} &
\colhead {$m_B$}}
\startdata
B & 3 & 0.644 $\pm$ 0.013 & -1.961 $\pm$ 0.015 & 18.377 $\pm$ 0.005 & 20.936 $\pm$ 0.030 \nl
R & 2 & 0.697 $\pm$ 0.009 & -1.979 $\pm$ 0.002 & 18.086 $\pm$ 0.009 & 19.948 $\pm$ 0.008 \nl
I & 3 & 0.698 $\pm$ 0.023 & -1.972 $\pm$ 0.006 & 17.664 $\pm$ 0.009 & 19.517 $\pm$ 0.017 \nl
\enddata
\tablecomments{$N_{im}$ is the number of images used in the analysis.  Error
bars are $\left(\sigma^2/N_{im} \right)^{1/2}$ errors from the observed 
dispersion between the images.  Relative coordinates are for component B with respect
to component A.}
\end{deluxetable}

\clearpage



 

\makeatletter
\def\jnl@aj{AJ}
\ifx\revtex@jnl\jnl@aj\let\tablebreak=\nl\fi
\makeatother


\begin{deluxetable}{crrccc}
\tablecaption{Relative Astrometry and Absolute Photometry for Nearby Reference Objects
\label{TABLE3}}
\tablenum{3}
\tablewidth{0pt}
\tablehead{
\colhead {Object} &
\colhead {$\Delta \alpha (^s)$} &
\colhead {$\Delta \delta ('')$} &
\colhead {$m_B$} &
\colhead {$m_R$} &
\colhead {$m_I$}}
\startdata
1 & -15.079& 120.82& 20.593 $\pm$ 0.268 & 18.345 $\pm$ 0.023 & 16.942 $\pm$ 0.010 \nl 
2 &  -8.816& 108.12& 21.592 $\pm$ 0.104 & 18.586 $\pm$ 0.032 & 17.081 $\pm$ 0.021 \nl 
3 &  -7.252&  17.20& 19.751 $\pm$ 0.057 & 18.041 $\pm$ 0.022 & 17.472 $\pm$ 0.004 \nl 
4 &  -3.712&  52.20& 20.522 $\pm$ 0.050 & 19.020 $\pm$ 0.049 & 18.538 $\pm$ 0.029 \nl 
5 &   0.000&   0.00& 18.456 $\pm$ 0.015 & 15.923 $\pm$ 0.007 & 15.025 $\pm$ 0.003 \nl 
6 &   1.093&  72.13& 20.706 $\pm$ 0.146 & 18.992 $\pm$ 0.081 & 18.474 $\pm$ 0.091 \nl 
7 &   9.281& -40.45& 17.955 $\pm$ 0.018 & 16.342 $\pm$ 0.002 & 15.816 $\pm$ 0.005 \nl 
8 &  10.718& -46.98& 19.366 $\pm$ 0.040 & 17.034 $\pm$ 0.009 & 16.308 $\pm$ 0.002 \nl 
\enddata
\tablecomments{Magnitudes are from 9\farcs6 diameter aperture photometry from the Nov. 1998
CTIO data.  Object numbers correspond to the labels shown in Figure 1.
Reported error bars are $\left(\sigma ^2/N_{im}\right)^{1/2}$ errors from the 
observed dispersion between the images.}
\end{deluxetable}


\clearpage



 

\makeatletter
\def\jnl@aj{AJ}
\ifx\revtex@jnl\jnl@aj\let\tablebreak=\nl\fi
\makeatother


\begin{deluxetable}{lcccc}
\tablecaption{Redshift Analysis for CTQ 839 A,B
\label{TABLE4}}
\tablenum{4}
\tablewidth{0pt}
\tablehead{  & \multicolumn{2}{c}{A} & \multicolumn{2}{c}{B} \nl
 & \multicolumn{2}{l}{------------------------} &
   \multicolumn{2}{l}{------------------------} \nl
Feature & $\lambda_{obs}$ & $z$ & $\lambda_{obs}$ & $z$ }
\startdata
Ly $\alpha$ & 3943.8 & 2.244(1)       & 3950.1 &  2.249(1) \nl
\ion{N}{5}  & 4020.7 & 2.240(1)       & 4018.1 &  2.238(4) \nl
\ion{O}{1}  & 4228.4 & 2.244(2)       &   ---  &   ---     \nl
\ion{C}{4}  & 5019.4 & 2.239(1)       & 5020.8 &  2.240(4) \nl
\ion{C}{3]} & 6187.9 & 2.242(2)       &   ---  &   ---     \nl
\enddata
\tablecomments{Numbers in parenthesis are $\pm 1 \sigma$ uncertainties in
the last digit of the quoted redshift.}
\end{deluxetable}

\clearpage

\begin{figure}[h]
\vspace{7.0 truein}
\includegraphics{NM.fig1.ps}
\end{figure}
\clearpage

\begin{figure}[h]
\vspace{7.0 truein}
\includegraphics{NM.fig2.ps}
\end{figure}

\clearpage

\thispagestyle{empty}
\begin{figure}[h]
\vspace{7.0 truein}
\includegraphics{NM.fig3.ps}
\end{figure}

\clearpage

\thispagestyle{empty}
\begin{figure}[h]
\vspace{7.0 truein}
\includegraphics{NM.fig4.ps}
\end{figure}

\clearpage

\thispagestyle{empty}
\begin{figure}[h]
\vspace{7.0 truein}
\includegraphics{NM.fig5.ps}
\end{figure}



\clearpage


\begin{references}

\reference{bald1984} Baldwin, J. A., \& Stone, R. P. S. 1984, \mnras, 206, 241

\reference{barn1999} Barnes, J. E. 1999, in AIP Conf. Proc. {\it After 
 the Dark Ages: When Galaxies Were Young}, eds. S. S. Holt \& E. P. Smith 
 (New York: AIP), 470, 191

\reference{clae1996} Claeskens, J. -F., Surdej, J., \& Remy, M. 1996, 
 \aap, 305, L9

\reference{cole1980} Coleman, G. D., Wu, C. -C., \& Weedman, D. W. 1980,
 \apjs, 43, 393

\reference{cond1998} Condon, J. J., Cotton, W. D., Greisen, E. W., Yin, Q. F.,
 Perley, R. A., Taylor, G. B., \& Broderick, J. J. 1998, \aj, 115, 1693

\reference{fuku1995} Fukugita, M., Shimasaku, K., \& Ichikawa, T. 1995, \pasp,
 107, 945

\reference{kays1986} Kayser, R., Refsdal, S., \& Stabell, R. 1986, \aap, 166, 
 36

\reference{keet1998} Keeton, C. R, Kochanek, C. S., \& Falco, E. E. 1998, \apj,
 509, 561 

\reference{koch1992} Kochanek, C. S. 1992, \apj, 384, 1

\reference{koch1996} Kochanek, C. S. 1996, \apj, 466, 638

\reference{koch1998} Kochanek, C. S., Falco, E. E., Impey, C. D., Leh\'ar, 
 J., McLeod, B. A., \& Rix, H., -W. 1998, in AIP Conf. Proc. 470, After
 the Dark Ages: When galaxies Were Young, ed. S. S. Holt \& E. P. Smith 
 (New York: AIP), 163

\reference{koch1999} Kochanek, C. S., Falco, E. E., \& Mu\~noz, J. A. 1999,
 \apj, 510, 590

\reference{land1992} Landolt, A. U. 1992, \aj, 104, 340

\reference{lill1997} Lilly, Simon.  1997, Private Communication

\reference{maza1995} Maza, J., Wischnjewsky, M., \& Antezana, R., 1996, 
 R.Mx.A.A., 32, 35

\reference{morg1999} Morgan, N. D., Dressler, A., Maza, J., Schechter, P. L.,
 \& Winn, J. N. 1999, \aj, 118, 1444

\reference{nara1998} Narayan, R., \& Bartlemann, M. 1998, in {\it Formation 
 of Structure in the Universe}, eds. A. Dekel \& J. Ostriker (Cambridge:
 Cambridge Univ. Press)

\reference{oste1993} Osterbrock, D. E. 1993, \apj, 404, 551

\reference{pers1992} Persson, S. E., West, S. C., Carr, D. M., 
 Sivaramakrishnan, A., \& Murphy, D. C. 1992, \pasp, 104, 204

\reference{pers1998} Persson, S. E., Murphy, D. C., Krzeminski, W., Roth, M., 
 \& Rieke, M. J. 1998, \aj, 116, 2475

\reference{refs1964} Refsdal, S., 1964, \mnras, 128, 307

\reference{sche1993a} Schechter, P. L., Mateo, M., \& Saha, A., 1993, 
 \pasp, 105, 1342

\reference{sche1993b} Schechter, P. L. \& Moore, C. B., 1993, \aj, 105, 1

\reference{sche1998} Schechter, P. L., Gregg, M. D., Becker, R. H., Helfand,
 D. J., \& White, R. L. 1998, \aj, 115, 1371

\reference{wist1999} Wisotzki, L., Christlieb, N., Liu, M. C., Maza, J.,
 Morgan, N. D., \& Schechter, P. L. 1999, \aap, 348, L41 

\end{references}
\end{document}